\documentclass[runningheads]{llncs}

\usepackage{graphicx}

\usepackage{multirow}
\usepackage{bm}
\usepackage{amsmath}
\usepackage{xcolor}
\usepackage{hyperref}
\usepackage{graphicx}

\begin{document}
\title{Stay on Topic, Please: Aligning User Comments to the Content of a News Article}
\titlerunning{Stay on Topic, Please: ACAP}

\author{Jumanah Alshehri \inst{1}$^{\star}$ \href{https://orcid.org/0000-0002-0077-7173}{\includegraphics[scale=0.5]{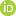}}
\and Marija Stanojevic \inst{1} \thanks{First two authors have contributed equally.} \href{https://orcid.org/0000-0001-8227-6577}{\includegraphics[scale=0.5]{figs/orcid_16x16.png}}
\and Eduard Dragut\inst{1}\href{https://orcid.org/0000-0002-3103-054X}{\includegraphics[scale=0.5]{figs/orcid_16x16.png}} 
\and
\newline Zoran Obradovic\inst{1}\href{https://orcid.org/0000-0002-2051-0142}{\includegraphics[scale=0.5]{figs/orcid_16x16.png}}}

\authorrunning{J. Alshehri \, M. Stanojevic \, E. Dragut \, Z. Obradovic}

\institute{Center for Data Analytics and Biomedical Informatics, Temple University, Philadelphia, PA, USA\\
\email{\{jumanah.alshehri,marija.stanojevic,edragut,zoran.obradovic\}@temple.edu}}

\maketitle 

\vspace{-10pt}
\begin{abstract}
Social scientists have shown that up to $50\%$ of the comments posted to a news article have no relation to its journalistic content. 
In this study we propose a classification algorithm to categorize user comments posted to a news article based on their alignment to its content. The alignment seeks to match user comments to an article based on similarity of content, entities in discussion, and topics. 
We propose a BERTAC, BERT-based approach that learns jointly article-comment embeddings and infers the relevance class of comments. We introduce an ordinal classification loss that penalizes the difference between the predicted and true labels. We conduct a thorough study to show influence of the proposed loss on the learning process. The results on five representative news outlets show that our approach can learn the comment class with up to $36\%$ average accuracy improvement comparing to the baselines, and up to $25\%$ comparing to the BA-BC. BA-BC is our approach that consists of two models aimed to capture dis-jointly the formal language of news articles and the informal language of comments. We also conduct a user study to evaluate human labeling performance to understand the difficulty of the classification task. The user agreement on comment-article alignment is ``moderate'' per Krippendorff’s alpha score, which suggests that the classification task is difficult.

\vspace{-5pt}

 \vspace{-10pt}
\end{abstract}

%%%%%%%%%%%%%%%%%%%%%%%%%%%%%% INTRODUCTION %%%%%%%%%%%%%%%%%%%%%%%%%%%%%%%
\section{Introduction}
\label{sec:introduction}
\vspace{-5pt}
The study of user comments is essential for social scientists, policymakers, and journalists since virtual discussions offer an insight into the public opinion and reaction to the daily news stream. In 2020, people shifted even more toward online discussions due to COVID-19. 
Many survey-based studies tried to understand the users' behavior by characterizing and categorizing comments in online news \cite{ruiz2011public,mishne2006leave,ziegele2013conceptualizing,weber2014discussions}. A salient outcome of these studies is that $20\%$ to $50\%$ of users' comments are irrelevant to the content or topic of those articles since users drift from the original topic to irrelevant sub-discussions \cite{singer2009separate}. Our goal in this work is to understand commenting behavior, more precisely, to automatically identify the subset of comments, from the set of comments an article receives, that are pertinent to the content of the article. The challenge is multi-fold: e.g., comments tend to be terse, colloquial, often nonliterary, containing grammatical errors, misspellings, and punctuation misuse. Our premise is that users are inclined to write comments that diverge from the article topic to different extents, especially in lengthier discussions. This noise in the data affects downstream applications such as opinion mining.

Previous studies tried to remove the noise among comments by studying toxic comments \cite{hosseini2017deceiving,georgakopoulos2018convolutional}, topic drifting \cite{info:doi/10.2196/jmir.6297,10.1007/978-3-319-41754-7_18}, and understanding the quality of online news comments \cite{10.1007/978-3-030-15719-7_23,Diakopoulos:2011:TQD:1958824.1958844,gottipati_jiang_2012_finding}. From a natural language processing (NLP) perspective, this problem is a supervised classification task to separate relevant from irrelevant comments.

In this paper, we introduce the Article-Comment Alignment Problem (ACAP). We aim to define a set of article-comment relevance classes and propose a methodology to classify article-comments pairs automatically. ACAP is a challenging task, for example, consider the article \textit{``This is going to happen in the United States: Donald Trump calls for surveillance of Muslims and advocates waterboarding terror suspects after Brussels attack''} \footnote{Full article: https://dailym.ai/2Qz7RG9} from \textit{Daily Mail} and the comment \textit{``It's not Europe anymore. It's Eurabia. This should not be a news story anymore.''} Two human annotators rate the comment as \textit{Irrelevant}, while the third annotator rates it as \textit{Same Category}. The third annotator's label is the most appropriate, but choosing that category requires background knowledge on the political circumstances in Europe in 2016. In solving ACAP, we hypothesize the following: 1) It is possible to capture the extent of a connection and semantics between an article and its comments using globally pre-trained models, fine-tuned with local data. 2) Considering the natural order of labels during training will boost the algorithm learning process.

We test our hypotheses in the following practical scenarios: (1) limiting amount of labeled article-comment pairs (1K per dataset), (2) bounding the number of tokens from each document (article or comment), and (3) concomitantly working with formal text, in the form of news articles, and informal text, in the form of comments. The article-comment pairs are extracted from five online news outlets: \textit{Wall Street Journal} (WSJ), \textit{Fox News} (FN), \textit{Daily Mail} (DM), \textit{The Guardian} (TG), and \textit{Market Watch} (MW). This work makes the following contributions:
\vspace{-5pt}
\begin{enumerate}
    \item We introduce the Article-Comment Alignment Problem (ACAP) and analyze the hardness of ACAP using an agreement study on the classification of human annotators.
    \item We propose BERTAC, which jointly learns embedding representations for articles and their comments, to solve ACAP. We also propose BA-BC, which consists of two models on trained on articles and the other on comments,  which attempts to capture the difference in language style between them, formal versus informal. We compare it to several approaches, including BA-BC, and show its superior performance.
    \item We develop a novel ordinal classification loss for BERTAC that penalizes the difference between the predicted and true labels. The proposed loss exhibits similar performance to the original loss in terms of accuracy, however, it boosts the model performance when trained on high agreement examples.
    \item We conduct extensive empirical studies on articles and comments from $5$ representative news outlets.
\end{enumerate}

%%%%%%%%%%%%%%%%%%%%%%%%%%%%% Related Work %%%%%%%%%%%%%%%%%%%%%%%%%%%%%% 
\section{RELATED WORK}
\label{relatedwork}
\vspace{-5pt}
User comments are a powerful means to understand public opinion and reaction to emerging events. Many organizations invest in mining user comments to improve their decision making. News outlets and social platforms are recommending most relevant user posts to keep the attention of busy readers  \cite{weber2014discussions}. Many studies focus on mining the user opinion from social media \cite{bastos2018parametrizing,celli2016predicting,stanojevic2019surveying} and online news comments \cite{almoqbel2019understanding,doi:10.1080/17512786.2014.899758,YangDM20a,ZhangYZDM20}. Other works look into bias in the news, and its influence on user-generated content \cite{stanojevic2019biased,YangDM20c}. The main challenge in those studies is the unpredictable quality of user-generated content.

To solve this problem, a line of research focuses on comment drifting \cite{10.1007/978-3-319-41754-7_18,info:doi/10.2196/jmir.6297} by utilizing the temporal nature of comments. The older an article is, the more commentators it has, and the probability of exposure to topic drift is higher \cite{10.1007/978-3-030-15719-7_23}. This phenomenon influences the quality of comments and their relevance.

Another line of work \cite{10.1145/2556195.2556231,10.1145/3072591,DBLP:conf/www/SilSB11} investigates which part of an article a comment aligns with using statistical models, while other \cite{gottipati_jiang_2012_finding} use hand-crafted structural, lexical, syntactic, discourse, and relevance features as an input to the logistic regression. Even though their F1 score is in the range of $70-80\%$ and their analysis measures correlation of the attributes with the label, hand-crafting features for each problem is difficult and time-consuming.

The work most related to ours attempts to automatically classify paragraph-comment agreement \cite{10.1007/978-3-030-15719-7_23}. They labeled the data based on Likert scale categories \cite{likert1932technique}, which is criticized for introducing bias. For instance, a number of works show that user responses are significantly affected by the order and direction of the rating scale \cite{article:scale2,article:scale}.

Instead, we propose to use transformer pre-trained language approach \cite{DBLP:journals/corr/abs-1810-04805,bertweet}.
We also create a new ordinal classification loss. As shown in the experimental study, our approach performs significantly better than the baselines on ACAP. We work with three annotators. Their labels give us support data to study the difficulty of the problem. We show that the annotators exhibit only fair agreement, indicating that ACAP is a difficult problem even for human beings.

%%%%%%%%%%%%%%%%%%%%%%%%%%%%% Data %%%%%%%%%%%%%%%%%%%%%%%%%%%%%% 
\section{DATASETS}
\label{sec:dataset}
\vspace{-5pt}
News articles and their comments were collected between Oct. $2015$ and Feb. $2017$ from Google News \cite{WIREs}.
The dataset has over $19$K articles with $9$M comments (including replies). For this study, we chose five news outlets that are representative of the problem at hand. The dataset contains articles and comments with a broad range of lengths and with a different number of comments, representative of the kind of news articles encountered at most news outlets. This data allows us to test the behavior of the proposed models under varied settings. Table \ref{tab:table1} (A) shows the statistics of datasets.
\begin{table}
\caption{(A) Statistics by outlet. We randomly selected $1$K article-comments pairs from each outlet and labeled them. ALA is the articles' average length and ALC is average comments' length, measured by number of words. (B) Classes proportions for each dataset. Outlets are sorted based on total number of available articles per outlet.} \label{tab:table1} \vspace{-5pt}
\begin{minipage}{2.4in}
\begin{tabular}{|l|c|c|c|c|}
\cline{1-5}\cline{1-5}
\multirow{2}{*}{Outlet} & \multicolumn{4}{c|}{(A) Dataset Statistics } \\\cline{2-5}
 & \#Art. & \#Comm. & \hspace{2pt} ALA \hspace{2pt}& \hspace{2pt} ALC \hspace{2pt} \\
\cline{1-5}\cline{1-5}
FN & $0.3$K & $72$K &$250$ & $22$\\
TG & $1.6$K & $428$K &$797$ & $54$\\
MW & $1.7$K & $65$K &$512$ & $42$\\
WSJ & $3.6$K & $309$K & $164$ & $57$\\
DM & $10$K & $1,012$K &$487$ & $28$\\
\cline{1-5}\cline{1-5}
\end{tabular}

\end{minipage}
\begin{minipage}{2.4in}
\begin{tabular}{|c|c|c|c|}
\cline{1-4}\cline{1-4}
\multicolumn{4}{|c|}{(B) Classes proportion } \\\cline{1-4}
Relevant & Same Ent. & Same Cat. & Irrelevant\\
\cline{1-4}\cline{1-4}
$3$\% & $21$\%  & $29$\% & $47$\%\\
$5$\% & $39\%$ & $32$\% & $24$\%\\
$7$\% & $51$\% & $20$\%  & $22$\%\\
$8$\% & $25$\% & $34$\% & $33$\%\\
$15$\% & $17$\% & $20$\% & $48$\%\\
\cline{1-4}\cline{1-4}
\end{tabular}
\end{minipage}
\end{table}  \vspace{-10pt}

\subsection{Labeling} 
\label{sec:labeling}
\vspace{-5pt}
We discard all articles without comments. We randomly select $1$K article-comment pairs from each outlet. Then, annotators manually and independently label the pairs in four classes: Relevant, Same Entities, Same Category, and Irrelevant. 

Relevant class - the content of the comment discusses the same matter as the content of the article. Same Entities class - the comment is not directly relevant, however, it mentions the same main entities within the same scope (category) of the article. For example, the article talks about a Real Madrid - F.C. Barcelona game, mentioning Ronaldo's performance in the game, and the comment talks about Ronaldo's best goal in the Portuguese team. Same Category class - comment in this class is not discussing the article, but it falls into the same category as the article. For example, both comment and article are discussing politics. Irrelevant class - a comment is in this class if it does not belong to any other class.  

\begin{figure}
    \centering
    \includegraphics[width=1\textwidth]{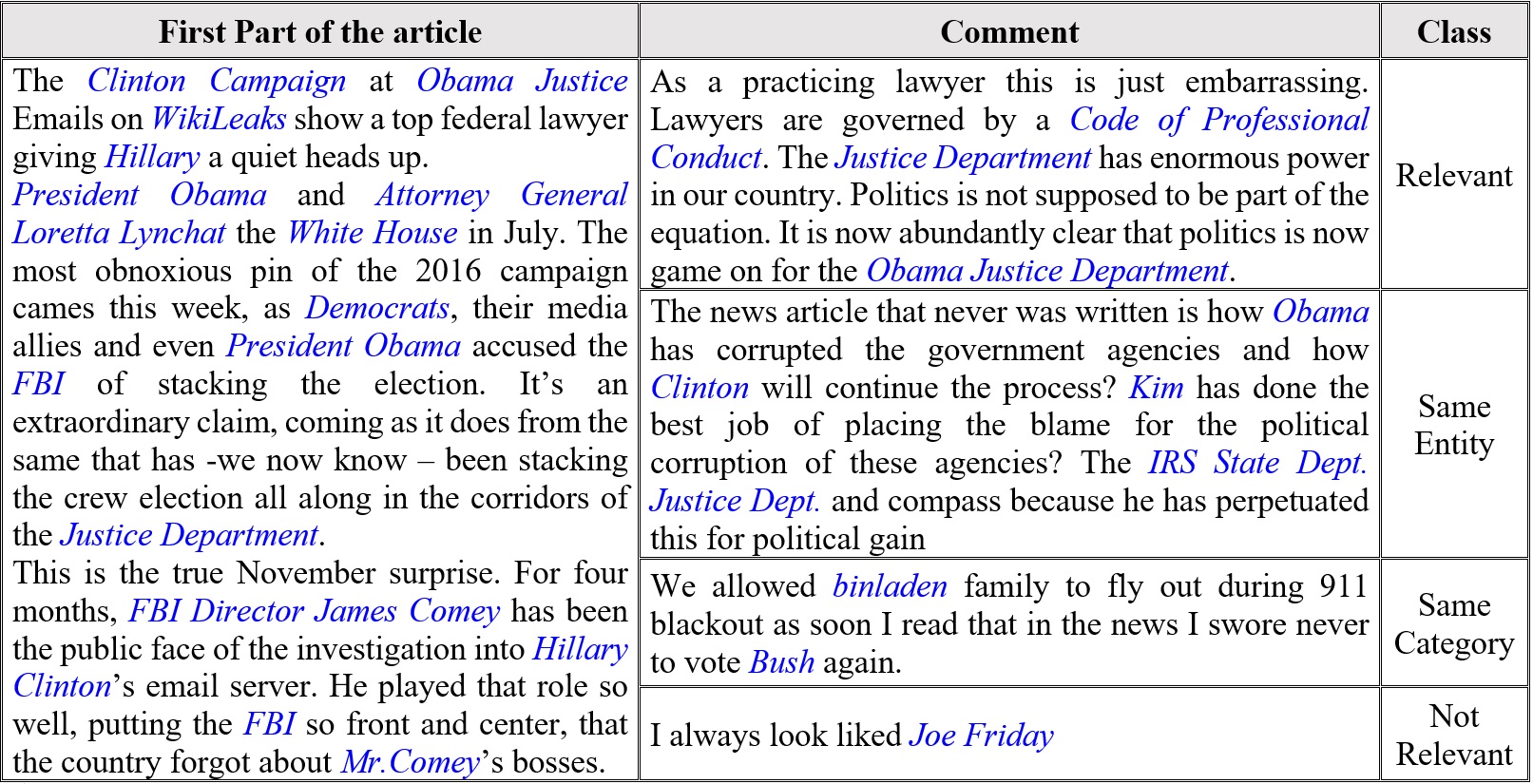} \vspace{-20pt}
    \caption{A labeling example from WSJ. Entities are colored. The article category is politics.}
    \label{fig:table2} \vspace{-10pt}
\end{figure}

Figure \ref{fig:table2} shows labeled examples of each class from our training data. The first column includes a part of the article; the second column has four comment examples, where each example represent a different class. The article in the table discusses Hillary Clinton's email story that came out before the U.S. election in 2016. The first comment is \textit{Relevant} since it discusses the same issue. The second comment does not discuss the main issue, however, it mentions some of the entities discussed in the article within the category of the article (politics). Hence, its class is \textit{Same Entities}. The third comment does not refer to any named entity from the article, but it goes into a political issue other than the ones mentioned in the article. Thus, its class is \textit{Same Category}. In the last comment, the user believes that he looks like \textit{Joe Friday}. This has no connection with the article, therefore, the comment is deemed \textit{Irrelevant}.

To obtain labeled instances, we asked three native English speakers, who were not involved in problem modeling nor solving it, to annotate the article-comments pairs. We provide them with the following pieces of information: 1) an article-comment pair without the surrounding context (i.e., the parent and child comments), and 2) the four label categories with an explanation and an example for each of them. Each article-comment pair receives 3 labels, one from each annotator. We assign the final label using an averaging aggregation scheme. We map Irrelevant, Same Category, Same Entity, and Relevant to  $0, 1, 2,$ and $3$, respectively. We average the (user) scores per pair and round to the nearest integer, which becomes the label of the pair. For example, a pair x-y receives the score $1, 1,$ and $2$ will have a label $1$, which corresponds to ``Same Category''. Table \ref{tab:table1} (B) shows the proportion of each class per outlet. We also binarize labels, by assigning 1 to any comment with the label Relevant, Same Category, and Same Entity and 0 to Irrelevant comments.\vspace{-5pt}

\subsection{User Agreement Study}
\label{sec:uas}
\vspace{-5pt}
Determining the relevance level of a comment to an article is not an easy task. Using agreement analysis, we measure human performance. We use Fleiss Kappa statistic and Krippendorff's alpha coefficient to compare the agreement between the three annotators.
\textit{Fleiss Kappa} statistic \cite{fleiss1971measuring} calculates agreement between multiple scorers on each category and then averages the scores over categories to produce the final statistics as in Equation \ref{eq:kappa}. The interpretation of Kappa value is as follow, [-1, 0) = poor agreement, [0, 0.20] = slight agreement, [0.21, 0.40] = fair agreement, [0.41, 0.60] = Moderate agreement, [0.61, 0.80] = substantial agreement, and [0.81, 1.00] = almost perfect agreement. Fleiss Kappa test is not suitable for ordinal labels, as it assigns the same error to a scorer who mislabeled the category ``Relevant'' with ``Same Entities'' and a scorer who mislabeled the category ``Relevant'' with ``Irrelevant''. \vspace{-5pt}
\begin{equation}
\small
\label{eq:kappa} 
FK = \frac{\sum_{i=1}^{N}\sum_{j=1}^{k}v^2_{ij} - Nm}{Nm(m-1)} 
\end{equation}
To account for the magnitude of error that a scorer makes, we use \textit{Krippendorff's alpha coefficient} \cite{krippendorff2011computing}, this statistic works by considering the distance between labels given by multiple scorers. It is calculated by subtracting the disagreement among assigned values normalized by the disagreement that is achieved if labels are assigned by chance. Formula is given in Equation \ref{eq:alpha}. In this equation, $\alpha \in [0, 1]$, where 0 indicates random scoring, and 1 indicates a perfect correspondence between annotators.
\begin{equation}
\small
\label{eq:alpha}
    \alpha = 1 - \frac{(n-1)\sum_{i}\sum_{j}o_{ij} * \delta^2_{ij}}{\sum_{i}\sum_{j}v_i * v_j * \delta^2_{ij}}
\end{equation}

Table \ref{tab:table2} gives the agreement scores for the three annotators for each of the five datasets. We note that labeling article-comment pairs in WSJ is the hardest task, with the smallest correspondence between the annotators. The raters' agreement is ``Fair'' for WSJ, TG, DM, and MW and borderline ``Moderate'' for FN, based on Fleiss Kappa statistic. The Krippendorff's alpha score\footnote{Calculted by http://dfreelon.org/utils/recalfront/recal-oir/ software using ordinal setting} is between 42\% and 66\% across the outlets. Both metrics indicate the difficulty humans have when assigning the category of comment in general.
\vspace{-5pt}
\begin{table}
\begin{center}
\caption{\label{tab:table2} Agreement analysis for annotators labels.} \vspace{-5pt}
\begin{tabular}{|l|c|c|c|c|c|}
\hline
Dataset & \hspace{3pt} WSJ \hspace{3pt} & \hspace{3pt} TG \hspace{3pt} & \hspace{3pt} DM \hspace{3pt} & \hspace{3pt} MW \hspace{3pt} & \hspace{3pt} FN \hspace{3pt} \\
\hline
 Fleiss Kappa & 0.22 & 0.36 & 0.37 & 0.40 & 0.45\\
 Krippendorff's $\alpha$ \hspace{3pt}& 0.42 & 0.60 & 0.61 & 0.64 & 0.66\\
\hline
\end{tabular}
\end{center} \vspace{-20pt}
\end{table}

%%%%%%%%%%%%%%%%%%%%%%%%%%%%% Methods %%%%%%%%%%%%%%%%%%%%%%%%%%%%%% 

\section{Methods}
\label{sec:methods}
\vspace{-5pt}
This section describes BERTAC, BA-BC models, and the ordinal classification loss.\vspace{-5pt}

\subsection{BERTAC Model - Joint Modeling of Article and Comments}
\label{sec:simbert}
\vspace{-5pt}
BERTAC leverages BERT$_{base}$ architecture, which allows us to learn more expressive embeddings for articles and comments. To solve ACAP we combine an article and its comment into a pair of segments and separate them with the special token [SEP]. Our goal is to make use of BERT's self-attention mechanism and bidirectional cross attention in an end-to-end fashion to encode the relevance between an article and its comments. One challenge in this setting is that of determining the length (in words) of the input segments that allow the deep network architecture to encode useful article-comment relations. We explore multiple input lengths for each dataset based on the average length of the articles shown in Table \ref{tab:table1}. \vspace{-5pt}

\subsection{BA-BC Model - Disjoint Modeling of Article and Comments}
\label{sec:bbweet-model}
\begin{figure}[t]
%\vspace{-10pt}
    \centering
    \includegraphics[width=0.5\textwidth]{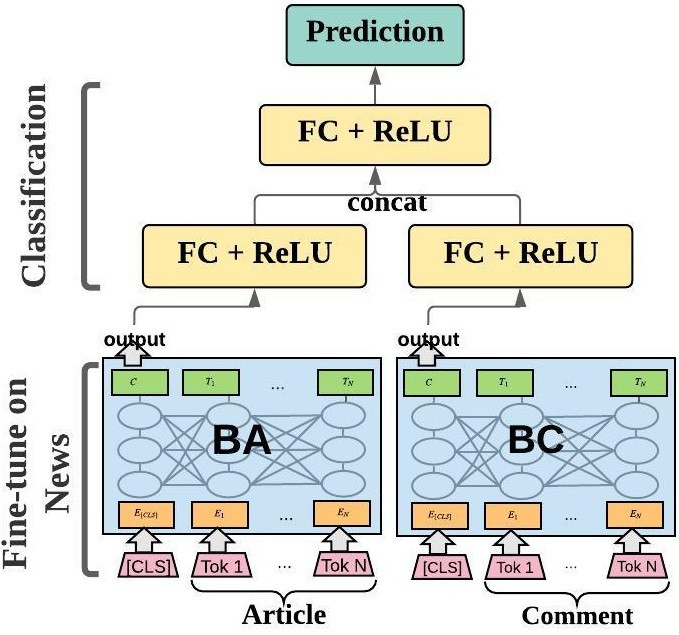} \vspace{-5pt}
    \caption{BA-BC Model for ACAP}
    \label{fig:bert-bertweet} \vspace{-10pt}
\end{figure}

\vspace{-5pt}
Our problem consists of two main parts. The first part is that of article representation, where the language is formal and usually formed of long sequences. The second part is the comment representation and alignment, where comments are often written in casual, informal language and consist of short sequences. We explore if a mixture of different pre-trained models can better solve ACAP. We call it BA-BC. Figure \ref{fig:bert-bertweet} shows the architecture of BA-BC. The model consists of two stages. The first stage \textit{Fine-tune on News} has two sides: (BA) is a BERT$_{base}$ architecture trained on articles and fine-tuned on articles; the second side (BC) is a BERTweet architecture trained on comments and fine-tuned on comments. BERTweet is a pre-trained model proposed by \cite{bertweet} and trained on $850 M$ English Tweets, the underlie architecture is RoBERTa \cite{liu2019roberta}. Their results show that BERTweet outperform RoBERTa$_{base}$ and XLM-R$_{base}$ \cite{xlm} in many tasks.

In the second stage, \textit{Classification} stage, the output from the first stage is fed into a fully-connected layer with a ReLU non-linearity. To get full advantage of pre-trained models, we designed two versions of BA-BC. The first is called \textit{BC-BA-Emb}, where the output from the \textit{Fine-tune on News} stage contains the Embeddings created by both sides without seeing any training examples from our datasets. The second model, called \textit{BC-BA-Fine-tune}, is additionally fine-tuned in the \textit{Fine-tune on News} stage. Left-side (BA) is fine-tuned on articles and labels and right-side (BC) is fine-tuned on comments and labels. Then, last hidden state of fine-tuned parts is sent to the second, \textit{Classification} stage. \vspace{-5pt}

\subsection{Ordinal Classification Loss}
\vspace{-5pt}
We introduce the ordinal classification loss, which accounts for the distance between the predicted class and the actual class. Here, we multiply the loss for each example with a weight that is calculated according to equation \ref{eq:ordinal_loss}, where $k=4$ (number of classes), $y_i$ is the actual label and $\bar{y_i}$ is the predicted label of the example.
\vspace{-5pt}
\begin{equation}
\small
\label{eq:ordinal_loss} 
weight = 1 + \frac{|\bar{y_i} - y_i|}{k - 1}
\end{equation}

If the algorithm chooses the right class, the weight is 1, so the loss is equal to original loss for correct prediction. If the model predicts a wrong category, the classification loss is multiplied by 2, 3, or 4 based on the distance between the real class and predicted class. The proposed loss depends on both the difference between the predicted class and the correct class, and the softmax error during predicting the actual category. We incorporated the proposed loss into BERTAC and compare it to original loss in Section \ref{sec:ordinalloss}.
\vspace{-5pt}

%%%%%%%%%%%%%%%%%%%%%%%%%%%%% Experimental Setup %%%%%%%%%%%%%%%%%%%%%%%%%%%%%% 
\section{EXPERIMENTAL SETUP}
\label{sec:expsetup}
\vspace{-5pt}In this section, we present the experimental setup all model and evaluation metrics used in this paper. \vspace{-5pt}

\subsection{Environment}
\label{sec:env}
\vspace{-5pt}
We run BERTAC, BA-BC, and Siamese LSTM models on four large nodes with 512 GB of DDR4 2400MHz RAM. Each of those machines has two sockets with Intel Xeon E5-2667 v4 3.2GHz processors, and every node contains two NVIDIA Tesla P100 PCIe 12GB GPUs and SSDs as local hard drives. Since doc2vec experiments are less computationally expensive, we run them on a 64-bit processor, Intel Core i7-6700 CPU @ 2.60 GHz with four cores and 16.0 GB RAM. \vspace{-5pt}

\subsection{Model evaluation}
\label{sec:modeleval}
\vspace{-5pt}
To evaluate the performance of the models, we use both simple and weighted accuracy since our labels are ordinal. With simple accuracy metric, which is calculated as the percent of correct predictions, predicting 0 or 2 for label 3 are counted as equal mistakes. Instead, we use weighted accuracy to calculate error by summing the absolute difference between predicted class $\widetilde{s_{i}}$ and ground truth $\bar{s_{i}}$. Model's error on a dataset is calculated by dividing that error by the number of examples and max difference $D$ between predicted classes. This constant is 3 in the given multi-class settings. The following formula computes the weighted accuracy, where $m$ is the number of examples:
\vspace{-5pt}
\begin{equation}
\small
\label{eq:evaluation}
WACC = 1 - \frac{\sum_{i=1}^{m}|\widetilde{s_{i}} - \bar{s_{i}}|}{mD}
\end{equation} \vspace{-10pt}

For all supervised models, experiments are repeated five times on different randomized split of labeled data. The dataset is split into 70:20:10 ratio for training, testing, and cross-validation, respectively. We report the mean and standard deviation. \vspace{-5pt}

\subsection{Comparison models}
\label{sec:comparisonmodels}
\vspace{-5pt}
A key challenge in solving ACAP is to establish a similarity of article-comment pair that is indicative of the relevance of the comment to the message of the article. We seek models that can learn long text representations using context and capitalize on the sequential nature of words in a comment. Besides, a model has to be able to embed two types of sequences: articles, which follow formal language, and comments, which may follow colloquial language.

\textit{Doc2vec} is the first baseline. Since it is unsupervised we use all data, comments, and articles available in each outlet to learn the documents embedding. We learn separate embeddings per outlet to account for the linguistic style accommodations and other biases across outlets \cite{Mahendiran2014DiscoveringEP}. Then, we calculate the cosine similarity score for all labeled article-comment pairs and assign a class for each pair based on the rules written below. $A$ represents articles, and $C$ represents comments. We experiment with thresholds in increments of 0.1. The ones used below give the best performance:
\[ \scriptsize
f(A_i, C_i) = 
     \begin{cases}
      \text{0,} &\text{if }cos(A_i, C_i)\le0.4\\
      \text{1,} &\text{otherwise, if }cos(A_i, C_i)\le0.6 \\
      \text{2,} &\text{otherwise, if }cos(A_i, C_i)\le0.8\\
      \text{3,} &\text{otherwise.} \\ 
     \end{cases}
\]
 
The second baseline is \textit{Siamese LSTM}, which consists of two LSTM that learn representations of articles and comments in separate modules. On top of these modules, there is a joint loss computation module, which computes the similarity between vectors and uses a dense layer to predict the label. Following \cite{mueller2016siamese}, we use the Manhattan distance to calculate the similarity for an article-comment pair.
We utilize a sparse categorical cross-entropy with a softmax activation function. 

\textit{BA-BC:} To produce the vector representation of articles and comments two main steps are required, embeddings and classifications. In \textit{BA-BC-Emb} model each side learns the embeddings by fine-tuning on articles and comments, respectively, in an unsupervised manner (without using labels). However, for \textit{BA-BC-Fine-tune}, each side is fine-tuned on articles and comments in a supervised way (using their associated labels). Later on, the vector representation of the last hidden layer is injected into the classification stage. The \textit{Fine-tune on News} stage is repeated for $3$ epochs and the final representation is fed into the \textit{Classification} stage, in which cross-entropy with a softmax activation function is applied as a loss. The classification stage is repeated for $300$ epochs.

\textit{BERTAC:} We leverage BERT$_{base}$ where it consists of $12$ layers, hidden layer size is $768$, number of self-attention heads is $12$, and the total number of parameters is $110$M. BERTAC is trained in two modes, cased and uncased, where letter casing is considered in the first while all letters are converted to small letters in the later. We trained the model for $6$ epochs.

%%%%%%%%%%%%%%%%%%%%%%%%% Results and Discussion %%%%%%%%%%%%%%%%%%%%%%%%%% 
\section{RESULTS AND DISCUSSION}
\label{sec:resultsanddiscussion}
\vspace{-5pt}
We study the complexity of this problem. In addition, we thoroughly study the multi-class datasets by employing and analyzing multiple models and evaluation measures to understand their behavior and identify the ones that better capture the semantic between an article and its comments. We also analyze the effect of the proposed Ordinal Classification Loss.
\vspace{-5pt}
\subsection{Binary versus Multiclass ACAP}
\label{sec:bvsm}
\vspace{-5pt}
To characterize the complexity of our problem we compare a binary dataset and multiclass dataset using BERTAC. In Table \ref{tab:binary_mlticlass_rsults}, we can see that the model maximal performance is around $92\%$ when the problem is binary. The accuracy drops between $13\% - 23\%$ when we have 4 classes. Even though having multiple classes helps people understand the relationship between an article and its comments better, it becomes harder for a model to capture the semantics and knowledge that is needed to distinguish some labels.
\vspace{-5pt}
%%%%%%%%%%%%%%%%% BINARY COMPARISONS
\begin{table}[ht]
\centering
\caption{\label{tab:binary_mlticlass_rsults} 
Test accuracy (in \%) for BERTAC. The \textit{B} row represents binary dataset and \textit{M} row represents multiclass dataset. The average test accuracy of 5 experiments is reported with standard deviation.}\vspace{-5pt}
\begin{tabular}{|c|c|c|c|c|c|c|}
\hline
Model & Dataset & FN & TG & MW & WSJ & DM \\\hline
\multirow{2}{*}{BERTAC} & B & $88.30 (1.42)$ & $92.45 (1.45)$ & $88.64 (2.50)$ & $85.07 (0.97)$ &  $90.46 (1.75)$  \\\cline{2-7}
 & M & $75.60 (1.81)$ & $74.58 (6.49)$ & $75.26 (4.52)$ & $63.17 (2.44)$ &  $67.36 (3.46)$   \\\hline
\end{tabular} \vspace{-10pt}
\end{table}

\subsection{Models Comparisons on Multiclass ACAP }
\label{sec:modelscomparisons}
\begin{figure}
    \centering
    \includegraphics[width=1\textwidth]{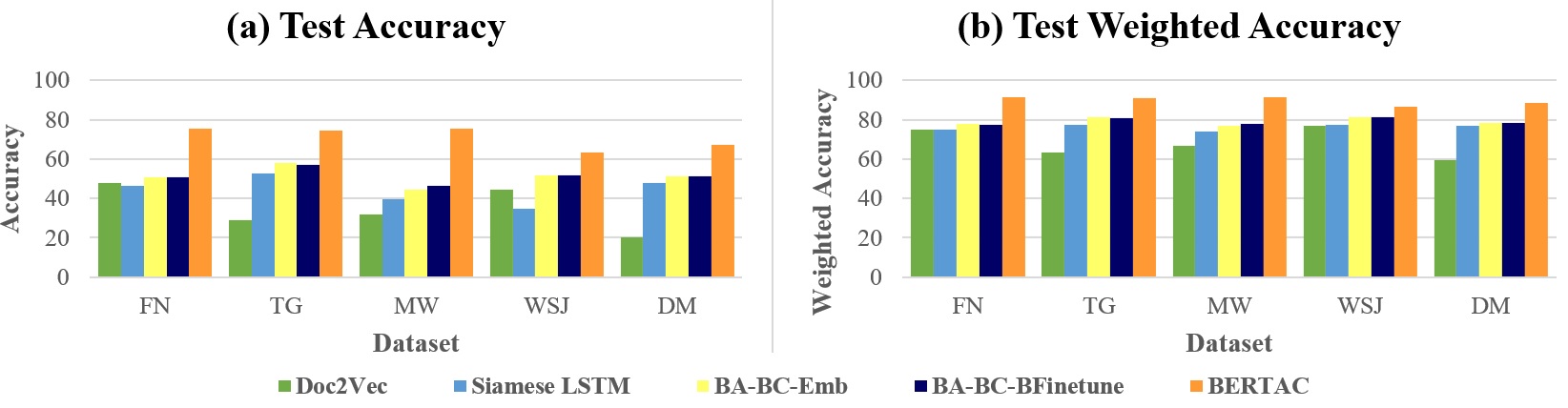} \vspace{-10pt}
    \caption{The average test accuracy of 5 experiments (in \%) for all models. (a) shows the accuracy results and (b) shows the weighted accuracy given by Equation \ref{eq:evaluation}.}
    \label{fig:test_results} \vspace{-20pt}
\end{figure}

\vspace{-5pt}
As shown in Figure \ref{fig:test_results}, performance of Doc2Vec is the worst among all models, despite training on much larger corpus of unsupervised data. Siamese LSTM accuracy exceeds Doc2Vec with a boost between $1\%$- $27\%$. BA-BC-Emb and BA-BC-Finetune outperform Siamese LSTM, the current SOTA for this problem. Both have an increase of $4\%$- $17\%$ in accuracy and $2\%$- $4\%$ in weighted accuracy over Siamese LSTM. When we compare BA-BC-Emb and BA-BC-Finetune we observe that they achieve similar performance. BERTAC however outperforms in both metrics all other models across all datasets when trained with the original loss function. In addition, we experiment with increasing the number of training points by merging the datasets. We note that increasing training examples does not improve any of the proposed models over BERTAC, which aligns with our hypothesis that BERTAC can outperform other models using only a small number of training examples.

Our experimental design is such that the number of articles varies between $300$ and $10,000$ across outlets as shown in Table \ref{tab:table1}. However, we label a fixed number of random outlet-comments. Therefore, the chance of selecting multiple comments for a single article is much larger at FN, TG and MW compared to WSJ and DM. A higher average accuracy is obtained on FN, TG and MW than on WSJ and DM. This suggests that when the model is trained on the same article with different comments and labels it can learn the pattern and predict the correct labels more accurately. 
\vspace{-5pt}
%%%%%%%%%%%%%%%%%%%%%% MULTI CLASS COMPARISON FIGURE 

\subsection{Weighted versus Un-Weighted Accuracy}
\label{sec:weightedacc}
\vspace{-5pt}
Comparing the outcomes of the algorithms on the weighted and unweighted accuracy, we note that their relative performance is unchanged: the model that has the lowest unweighted accuracy has the smallest weighted accuracy performance as well. The same relation stands for the highest results. There are a couple of explanations for this outcome. First, it suggests that, most of the wrongly classified instances are mixed with its neighboring classes. Second, it indicates a proportional number of stronger misclassifications (where the distance between the actual and predicted category is larger than 1) is present across news outlets.

The analysis of weighted accuracy helps to gain additional insight into the models and the problem hardness. For instance, by comparing the weighted and unweighted accuracy scores, we get a better idea of how well a model learns, since it accounts for the strength of the error, penalizing more the mistakes on harder examples. We observe a large gap between accuracy and weighted accuracy results, where the error is between 2 or 3 times smaller in most cases. This indicates that when a model misclassifies an example, often, the model predicts one of the neighboring classes to the correct class. For example, in \textit{WSJ}, \textit{Doc2Vec}'s accuracy = $44.41\%$ which is higher than of \textit{Siamese LSTM}'s accuracy = $34.81\%$. However, the weighted accuracy shows the opposite: \textit{Doc2Vec}'s accuracy = $76.88\%$ and \textit{Siamese LSTM}'s accuracy = $77.60\%$. This indicates that \textit{Siamese LSTM} is able to  understand the problem better and address the natural order of the classes during training.

\vspace{-5pt}
\subsection{Ordinal Classification Loss}
\label{sec:ordinalloss}
\vspace{-5pt}
We designed this experiment to investigate the effect of the \textit{ordinal loss} on BERTAC. We hypothesize that BERTAC trained with the proposed \textit{ordinal loss} will outperform the original loss. 

\begin{figure}[ht]
    \centering
    \includegraphics[width=1\textwidth]{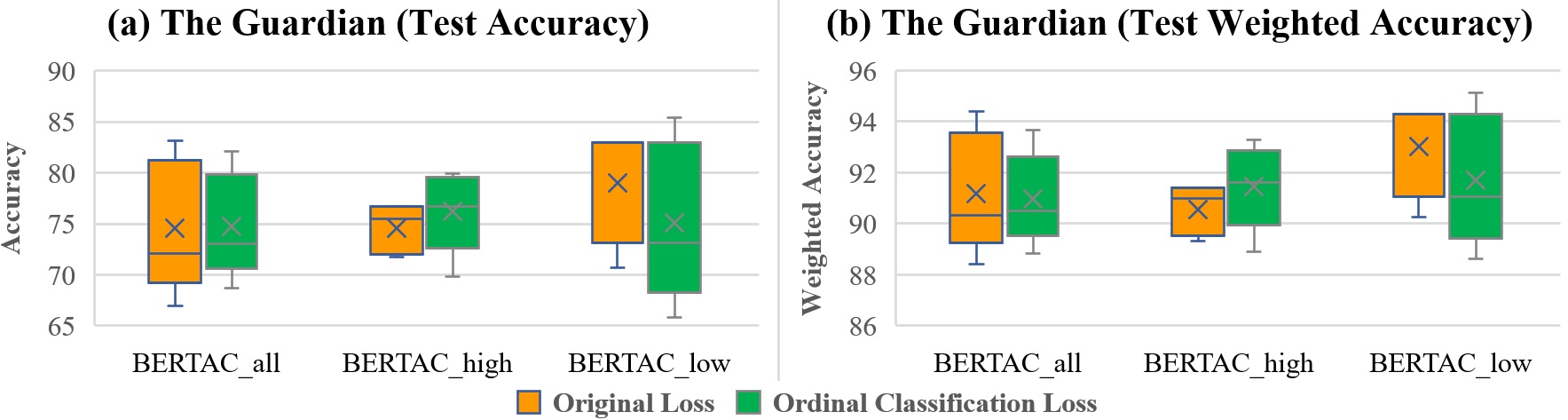}\vspace{-5pt}
    \caption{The Guardian average test accuracy of 5 experiments (a) show the accuracy results and (b) weighted accuracy. The subscript beside BERTAC indicates the experiment type, where $all$= trained on all labeled examples were used, $high$= trained on examples with high agreement score, and $low$= trained on examples with low agreement score.}
    \label{fig:tg_org_vs_ord}
\end{figure}

We find that \textit{proposed ordinal loss} has no significant advantage compared to \textit{original loss}, where both losses have similar performance. To better understand this problem we investigate those instances where the annotators highly agree with each other in the labeling task: $\sigma$ between the annotators's labels is either $0$, which means that they all agree, or $0.5$, which means that only one annotator disagree, with difference of $1$ and this does not affect the final label after aggregating the annotators labels. We call this the \textit{high agreement experiment} indicated by BERTAC$_{high}$ in Figure \ref{fig:tg_org_vs_ord}. On the other hand, the \textit{low agreement experiment} indicated by BERTAC$_{low}$, which contains only examples where $\sigma$ between the annotators' labels, is higher than $0.5$. We find that for some datasets BERTAC$_{low}$ accuracy is slightly higher than BERTAC$_{all}$ and BERTAC$_{high}$. However, looking into the high $\sigma$ we can see that the model is not consistent compared to BERTAC$_{high}$, and the number of examples are much fewer that BERTAC$_{all}$. Analyzing the original and ordinal losses for BERTAC$_{high}$, where annotators highly agree with each other, we find that the proposed ordinal loss is higher but not significantly. The improvement in accuracy is between $1\%-5\%$ and $1\%-3\%$ in the weighted accuracy. However, if we study the behavior across the models, we can see that ordinal loss behaves somehow differently across experiments. For examples in Figure \ref{fig:tg_org_vs_ord}, we can see that both BERTAC$_{high}$ and BERTAC$_{low}$ agree that ordinal loss is equal or better than original, where BERTAC$_{all}$ disagree. This brings the following question: \textit{if the model were capable to vote for the best possible prediction from different model would this improve results?}

\vspace{-5pt}
 \vspace{-5pt}

To answer the previous question, we calculate the average vote prediction from different models in order to obtain the best prediction. We consider the predictions from BERTAC uncased trained with ordinal loss and original loss, and BERTAC cased trained ordinal loss. Table \ref{tab:ord_vs_vote} shows that the voting system improves the results with respect to accuracy and standard division.

\vspace{-5pt}
\begin{table} \vspace{-5pt}
\small 
\centering 
\caption{\label{tab:ord_vs_vote} 
Accuracy results in \% for BERTAC trained with ordinal loss (BERTAC$_{ord}$) and BERTAC trained with different settings and losses (BERTAC$_{vote}$).} \vspace{-5pt}
\begin{tabular}{|c|c|c|c|c|c|}
\hline
Model & FN & TG & MW & WSJ & DM \\\hline
 BERTAC$_{ord}$ & $75.08 (4.19)$ & $74.78 (5.15)$ &  $71.08 (3.47)$ & $64.45 (3.36)$  & $68.42 (1.49)$  \\\hline
 BERTAC$_{vote}$ & $\textbf{76.73 (2.15)}$ & $\textbf{76.00 (6.16)}$ & $\textbf{74.00 (3.40)}$ & $\textbf{64.00 (2.77)}$ & $\textbf{69.02 (1.87)}$ \\\hline
\end{tabular} \vspace{-5pt}
\end{table}

%%%%%%%%%%%%%%%%%%%%%%%%%%%%% conclusion %%%%%%%%%%%%%%%%%%%%%%%%%%%%%%
\vspace{-20pt}
\section{CONCLUSION}
\label{sec:conclusion}
\vspace{-5pt}
In this work, we define the article-comment alignment problem (ACAP) and propose an effective approach to predict the level of relatedness between a comment and an article. We compare Doc2Vec, Siamese LSTM, BA-BC, and BERTAC models and study the performance improvement across them. The results reported in this work show that a joint modeling of article-comments, i.e., BERTAC, is able to capture a deeper level of semantic relatedness between comments and news articles, and help predict better the relevance level of a comment to the content of an article than the current state-of-the-art and other proposed methods.

Even though accuracy values are close, detailed analysis shows that BERTAC trained with proposed ordinal loss perform better than BERTAC on the original BERT loss. With the proposed loss, we can identify common mistakes by annotators and potentially use them to improve the performance of downstream applications, which we will explore in the future. 

%%%%%%%%%%%%%%%%%%%%%%%%%%%%%%%%%%%%%%%%%%%%%%%%%%%%%%%%%%%%%%%%%%%%%%%
\section*{Acknowledgements}
\label{sec:Ack}
This research was supported in part by the NSF grant IIS-8142183. In addition, this research includes calculations carried out on HPC resources supported in part by the National Science Foundation through major research instrumentation grant number 1625061 and by the US Army Research Laboratory under contract number W911NF-16-2-0189.

% \bibliographystyle{splncs04}
% \bibliography{References}

\begin{thebibliography}{10}
\providecommand{\url}[1]{\texttt{#1}}
\providecommand{\urlprefix}{URL }
\providecommand{\doi}[1]{https://doi.org/#1}

\bibitem{almoqbel2019understanding}
Almoqbel, M.Y., Wohn, D.Y., Hayes, R.A., Cha, M.: Understanding facebook news
  post comment reading and reacting behavior through political extremism and
  cultural orientation. Computers in Human Behavior  \textbf{100},  118 -- 126
  (2019). \doi{https://doi.org/10.1016/j.chb.2019.06.006},
  \url{http://www.sciencedirect.com/science/article/pii/S0747563219302250}

\bibitem{bastos2018parametrizing}
Bastos, M., Mercea, D.: Parametrizing brexit: mapping twitter political space
  to parliamentary constituencies. Information, Communication \& Society
  \textbf{21}(7),  921--939 (2018). \doi{10.1080/1369118X.2018.1433224},
  \url{https://doi.org/10.1080/1369118X.2018.1433224}

\bibitem{celli2016predicting}
Celli, F., Stepanov, E.A., Poesio, M., Riccardi, G.: Predicting brexit:
  Classifying agreement is better than sentiment and pollsters. In: Nissim, M.,
  Patti, V., Plank, B. (eds.) Proceedings of the Workshop on Computational
  Modeling of People's Opinions, Personality, and Emotions in Social Media,
  PEOPLES@COLING 2016, Osaka, Japan, December 12, 2016. pp. 110--118. The
  {COLING} 2016 Organizing Committee (2016),
  \url{https://www.aclweb.org/anthology/W16-4312/}

\bibitem{xlm}
Conneau, A., Khandelwal, K., Goyal, N., Chaudhary, V., Wenzek, G., Guzm{\'a}n,
  F., Grave, E., Ott, M., Zettlemoyer, L., Stoyanov, V.: Unsupervised
  cross-lingual representation learning at scale. In: Proceedings of the 58th
  Annual Meeting of the Association for Computational Linguistics. pp.
  8440--8451. Association for Computational Linguistics, Online (Jul 2020).
  \doi{10.18653/v1/2020.acl-main.747},
  \url{https://www.aclweb.org/anthology/2020.acl-main.747}

\bibitem{10.1145/2556195.2556231}
Das, M.K., Bansal, T., Bhattacharyya, C.: Going beyond corr-lda for detecting
  specific comments on news {\&} blogs. In: Carterette, B., Diaz, F., Castillo,
  C., Metzler, D. (eds.) Seventh {ACM} International Conference on Web Search
  and Data Mining, {WSDM} 2014, New York, NY, USA, February 24-28, 2014. pp.
  483--492. {ACM} (2014). \doi{10.1145/2556195.2556231},
  \url{https://doi.org/10.1145/2556195.2556231}

\bibitem{DBLP:journals/corr/abs-1810-04805}
Devlin, J., Chang, M., Lee, K., Toutanova, K.: {BERT:} pre-training of deep
  bidirectional transformers for language understanding. In: Burstein, J.,
  Doran, C., Solorio, T. (eds.) Proceedings of the 2019 Conference of the North
  American Chapter of the Association for Computational Linguistics: Human
  Language Technologies, {NAACL-HLT} 2019, Minneapolis, MN, USA, June 2-7,
  2019, Volume 1 (Long and Short Papers). pp. 4171--4186. Association for
  Computational Linguistics (2019). \doi{10.18653/v1/n19-1423},
  \url{https://doi.org/10.18653/v1/n19-1423}

\bibitem{Diakopoulos:2011:TQD:1958824.1958844}
Diakopoulos, N., Naaman, M.: Towards quality discourse in online news comments.
  In: Hinds, P.J., Tang, J.C., Wang, J., Bardram, J.E., Ducheneaut, N. (eds.)
  Proceedings of the 2011 {ACM} Conference on Computer Supported Cooperative
  Work, {CSCW} 2011, Hangzhou, China, March 19-23, 2011. pp. 133--142. {ACM}
  (2011). \doi{10.1145/1958824.1958844},
  \url{https://doi.org/10.1145/1958824.1958844}

\bibitem{fleiss1971measuring}
Fleiss, J.L.: Measuring nominal scale agreement among many raters.
  Psychological Bulletin  \textbf{76}(5),  378–--382 (1971),
  \url{https://doi.org/10.1037/h0031619}

\bibitem{article:scale2}
Friedman, H., ~, T.A.: Rating the rating scales. Journal of Marketing
  Management. pp. 114--123 (1999), \url{https://ssrn.com/abstract=2333648}

\bibitem{georgakopoulos2018convolutional}
Georgakopoulos, S.V., Tasoulis, S.K., Vrahatis, A.G., Plagianakos, V.P.:
  Convolutional neural networks for toxic comment classification. In:
  Proceedings of the 10th Hellenic Conference on Artificial Intelligence,
  {SETN} 2018, Patras, Greece, July 09-12, 2018. pp. 35:1--35:6. {ACM} (2018).
  \doi{10.1145/3200947.3208069}, \url{https://doi.org/10.1145/3200947.3208069}

\bibitem{gottipati_jiang_2012_finding}
Gottipati, S., Jiang, J.: Finding thoughtful comments from social media. In:
  Kay, M., Boitet, C. (eds.) {COLING} 2012, 24th International Conference on
  Computational Linguistics, Proceedings of the Conference: Technical Papers,
  8-15 December 2012, Mumbai, India. pp. 995--1010. Indian Institute of
  Technology Bombay (2012), \url{https://www.aclweb.org/anthology/C12-1061/}

\bibitem{10.1007/978-3-319-41754-7_18}
Gr{\"{u}}tze, T., Krestel, R., Naumann, F.: Topic shifts in stackoverflow: Ask
  it like socrates. In: M{\'{e}}tais, E., Meziane, F., Saraee, M., Sugumaran,
  V., Vadera, S. (eds.) Natural Language Processing and Information Systems -
  21st International Conference on Applications of Natural Language to
  Information Systems, {NLDB} 2016, Salford, UK, June 22-24, 2016, Proceedings.
  Lecture Notes in Computer Science, vol.~9612, pp. 213--221. Springer (2016).
  \doi{10.1007/978-3-319-41754-7\_18},
  \url{https://doi.org/10.1007/978-3-319-41754-7\_18}

\bibitem{WIREs}
He, L., Han, C., Mukherjee, A., Obradovic, Z., Dragut, E.: On the dynamics of
  user engagement in news comment media. Wiley Interdiscip. Rev. Data Min.
  Knowl. Discov.  \textbf{10}(1) (2020). \doi{10.1002/widm.1342},
  \url{https://doi.org/10.1002/widm.1342}

\bibitem{doi:10.1080/17512786.2014.899758}
Hille, S., Bakker, P.: Engaging the social news user. Journalism Practice
  \textbf{8}(5),  563--572 (2014). \doi{10.1080/17512786.2014.899758},
  \url{https://doi.org/10.1080/17512786.2014.899758}

\bibitem{hosseini2017deceiving}
Hosseini, H., Kannan, S., Zhang, B., Poovendran, R.: Deceiving google's
  perspective {API} built for detecting toxic comments. CoRR
  \textbf{abs/1702.08138} (2017), \url{http://arxiv.org/abs/1702.08138}

\bibitem{10.1145/3072591}
Hou, L., Li, J., Li, X., Tang, J., Guo, X.: Learning to align comments to news
  topics. {ACM} Trans. Inf. Syst.  \textbf{36}(1),  9:1--9:31 (2017).
  \doi{10.1145/3072591}, \url{https://doi.org/10.1145/3072591}

\bibitem{krippendorff2011computing}
Krippendorff, K.: Computing krippendorff's alpha-reliability. Scholarly Commons
   (2011)

\bibitem{likert1932technique}
Likert, R.: A technique for the measurement of attitudes. Archives of
  Psychology  (1932)

\bibitem{liu2019roberta}
Liu, Y., Ott, M., Goyal, N., Du, J., Joshi, M., Chen, D., Levy, O., Lewis, M.,
  Zettlemoyer, L., Stoyanov, V.: Roberta: {A} robustly optimized {BERT}
  pretraining approach. CoRR  \textbf{abs/1907.11692} (2019),
  \url{http://arxiv.org/abs/1907.11692}

\bibitem{Mahendiran2014DiscoveringEP}
Mahendiran, A., Wang, W., Lira, J.S., Huang, B., Getoor, L., Mares, D.,
  Ramakrishnan, N.: Discovering evolving political vocabulary in social media.
  In: 2014 International Conference on Behavioral, Economic, and Socio-Cultural
  Computing (BESC2014). pp.~1--7 (2014). \doi{10.1109/BESC.2014.7059504}

\bibitem{mishne2006leave}
Mishne, G., Glance, N.: Leave a reply: An analysis of weblog comments. In: In
  Third annual workshop on the Weblogging ecosystem (2006)

\bibitem{mueller2016siamese}
Mueller, J., Thyagarajan, A.: Siamese recurrent architectures for learning
  sentence similarity. In: Schuurmans, D., Wellman, M.P. (eds.) Proceedings of
  the Thirtieth {AAAI} Conference on Artificial Intelligence, February 12-17,
  2016, Phoenix, Arizona, {USA}. pp. 2786--2792. {AAAI} Press (2016),
  \url{http://www.aaai.org/ocs/index.php/AAAI/AAAI16/paper/view/12195}

\bibitem{10.1007/978-3-030-15719-7_23}
Mullick, A., Ghosh, S., Dutt, R., Ghosh, A., Chakraborty, A.: Public sphere
  2.0: Targeted commenting in online news media. In: Advances in Information
  Retrieval. pp. 180--187. Springer International Publishing, Cham (2019)

\bibitem{bertweet}
Nguyen, D.Q., Vu, T., Nguyen, A.T.: {BERT}weet: A pre-trained language model
  for {E}nglish tweets. In: Proceedings of the 2020 Conference on Empirical
  Methods in Natural Language Processing: System Demonstrations. pp. 9--14.
  Association for Computational Linguistics, Online (Oct 2020).
  \doi{10.18653/v1/2020.emnlp-demos.2},
  \url{https://www.aclweb.org/anthology/2020.emnlp-demos.2}

\bibitem{info:doi/10.2196/jmir.6297}
Park, A., Hartzler, A., Huh, J., Hsieh, G., DavidMcDonald, Pratt, W.: ``how did
  we get here?'': Topic drift in online health discussions. J Med Internet Res
  \textbf{18}(11), ~e284 (Nov 2016). \doi{10.2196/jmir.6297},
  \url{http://www.jmir.org/2016/11/e284/}

\bibitem{ruiz2011public}
Ruiz, C., Domingo, D., Micó, J.L., Díaz-Noci, J., Meso, K., Masip, P.: Public
  sphere 2.0? the democratic qualities of citizen debates in online newspapers.
  The International Journal of Press/Politics  \textbf{16}(4),  463--487
  (2011). \doi{10.1177/1940161211415849},
  \url{https://doi.org/10.1177/1940161211415849}

\bibitem{DBLP:conf/www/SilSB11}
Sil, D.K., Sengamedu, S.H., Bhattacharyya, C.: Readalong: reading articles and
  comments together. In: Srinivasan, S., Ramamritham, K., Kumar, A., Ravindra,
  M.P., Bertino, E., Kumar, R. (eds.) Proceedings of the 20th International
  Conference on World Wide Web, {WWW} 2011, Hyderabad, India, March 28 - April
  1, 2011 (Companion Volume). pp. 125--126. {ACM} (2011).
  \doi{10.1145/1963192.1963256}, \url{https://doi.org/10.1145/1963192.1963256}

\bibitem{singer2009separate}
Singer, J.B.: Separate spaces: Discourse about the 2007 scottish elections on a
  national newspaper web site. The International Journal of Press/Politics
  \textbf{14}(4),  477--496 (2009). \doi{10.1177/1940161209336659},
  \url{https://doi.org/10.1177/1940161209336659}

\bibitem{stanojevic2019biased}
Stanojevic, M., Alshehri, J., Dragut, E., Obradovic, Z.: Biased news data
  influence on classifying social media posts. In: Proceedings of the Third
  International Workshop on Recent Trends in News Information Retrieval,
  co-located with 42nd International {ACM} Conference on Research and
  Development in Information Retrieval {(SIGIR} 2019), Paris, France, July 25,
  2019. {CEUR} Workshop Proceedings, vol.~2411, pp.~3--8. CEUR-WS.org (2019),
  \url{http://ceur-ws.org/Vol-2411/paper1.pdf}

\bibitem{stanojevic2019surveying}
Stanojevic, M., Alshehri, J., Obradovic, Z.: Surveying public opinion using
  label prediction on social media data. In: {ASONAM} '19: International
  Conference on Advances in Social Networks Analysis and Mining, Vancouver,
  British Columbia, Canada, 27-30 August, 2019. pp. 188--195. {ACM} (2019).
  \doi{10.1145/3341161.3342861}, \url{https://doi.org/10.1145/3341161.3342861}

\bibitem{weber2014discussions}
Weber, P.: Discussions in the comments section: Factors influencing
  participation and interactivity in online newspapers' reader comments. New
  Media Soc.  \textbf{16}(6),  941--957 (2014). \doi{10.1177/1461444813495165},
  \url{https://doi.org/10.1177/1461444813495165}

\bibitem{article:scale}
Yan, T., Keusch, F.: {The Effects of the Direction of Rating Scales on Survey
  Responses in a Telephone Survey}. Public Opinion Quarterly  \textbf{79}(1),
  145--165 (02 2015). \doi{10.1093/poq/nfu062},
  \url{https://doi.org/10.1093/poq/nfu062}

\bibitem{YangDM20a}
Yang, F., Dragut, E., Mukherjee, A.: Claim verification under positive
  unlabeled learning. In: ASONAM (2020)

\bibitem{YangDM20c}
Yang, F., Dragut, E., Mukherjee, A.: Predicting personal opinion on future
  events with fingerprints. In: COLING (Dec 2020)

\bibitem{ZhangYZDM20}
Zhang, Y., Yang, F., Zhang, Y., Dragut, E., Mukherjee, A.: Birds of a feather
  flock together: Satirical news detection via language model differentiation
  (2020)

\bibitem{ziegele2013conceptualizing}
Ziegele, M., Quiring, O.: Conceptualizing online discussion value: A
  multidimensional framework for analyzing user comments on mass-media
  websites. Annals of the International Communication Association
  \textbf{37}(1),  125--153 (2013). \doi{10.1080/23808985.2013.11679148},
  \url{https://doi.org/10.1080/23808985.2013.11679148}

\end{thebibliography}

\end{document}